\documentclass[11pt, a4paper]{article}

\usepackage[utf8]{inputenc}
\usepackage[T1]{fontenc}
\usepackage{amsmath}          
\usepackage{graphicx}         
\usepackage{hyperref}         
\usepackage[margin=1in]{geometry} 
\usepackage{authblk}          
\usepackage{array}     
\usepackage{booktabs}  
\usepackage{multirow} 
\usepackage{enumitem}
\usepackage{natbib} 
\usepackage{tabularx}
\setlist[itemize,1]{topsep=2pt, partopsep=0pt}

\title{Synthetic Founders: AI-Generated Social Simulations for Startup Validation Research in Computational Social Science}

\author[1]{Jorn K. Teutloff}
\affil[1]{Marymount University}

\date{} 

\begin{document}

\maketitle

\begin{abstract}
We present a comparative docking experiment that aligns human-subject interview data with large language model (LLM)–driven synthetic personas to evaluate fidelity, divergence, and blind spots in AI-enabled simulation. Fifteen early-stage startup founders were interviewed about their hopes and concerns regarding AI-powered validation, and the same protocol was replicated with AI-generated founder and investor personas. A structured thematic synthesis revealed four categories of outcomes: (1) Convergent themes - commitment-based demand signals, black-box trust barriers, and efficiency gains were consistently emphasized across both datasets; (2) Partial overlaps - founders worried about outliers being averaged away and the stress of real customer validation, while synthetic personas highlighted irrational blind spots and framed AI as a psychological buffer; (3) Human-only themes - relational and advocacy value from early customer engagement and skepticism toward moonshot markets; and (4) Synthetic-only themes - amplified false positives and trauma blind spots, where AI may overstate adoption potential by missing negative historical experiences.

We interpret this comparative framework as evidence that LLM-driven personas constitute a form of hybrid social simulation: more linguistically expressive and adaptable than traditional rule-based agents, yet bounded by the absence of lived history and relational consequence. Rather than replacing empirical studies, we argue they function as a complementary simulation category - capable of extending hypothesis space, accelerating exploratory validation, and clarifying the boundaries of cognitive realism in computational social science.
\end{abstract}


\section{Introduction}
The use of simulation as a lens for understanding complex social systems has long been central to computational social science \citep{GilbertTroitzsch2005, Epstein1999}. From early agent-based models of market dynamics to contemporary simulations of policy interventions, researchers have relied on artificial agents to approximate the decision-making processes of humans \citep{Axelrod1997, Tesfatsion2002}. Yet traditional agent-based models remain limited by their reliance on simplified rules and narrow behavioral assumptions \citep{Sun2006}. In parallel, the emergence of large language models (LLMs) has created new possibilities for constructing agents that generate contextually rich, linguistically coherent, and psychologically nuanced responses \citep{Park2023, Argyle2023}.

This paper explores one such application: the use of AI-generated synthetic personas as simulation agents in the domain of entrepreneurship. Specifically, we demonstrate how founder and investor personas, derived from LLMs, can be treated as computational actors in a simulation experiment designed to systematically align with and extend a human interview study. We frame this as a comparative validation study - a methodological docking experiment - positioning synthetic personas as a novel methodological extension of social simulation. Situated between traditional rule-based agents and empirical field data, these agents offer researchers a new tool for probing decision-making under conditions of uncertainty.

Agent-based modeling (ABM) has been a cornerstone of social simulation for decades, enabling researchers to explore emergent system behaviors from micro-level rules \citep{EpsteinAxtell1996, MacyWiller2002}. Yet the strength of ABM - its reliance on parsimonious assumptions - also limits its capacity to capture the linguistic richness, cognitive biases, and psychological nuance of real human decision-making \citep{ContePaolucci2014}. This is a challenge that has animated methodological debates regarding parsimony versus realism \citep{Epstein2008, Edmonds2005}, including work on validation against empirical data \citep{Squazzoni2012} and the integration of psychological mechanisms into ABMs \citep{Jager2017}. Our approach contributes to this debate by showing how docking experiments can assess whether richer, linguistically expressive agents genuinely add explanatory value without sacrificing methodological rigor. Simplified behavioral rules often underrepresent the irrationalities, “polite lies,” and coping strategies that characterize entrepreneurial behavior in practice. This gap has constrained the ability of simulation studies to faithfully represent high-stakes domains like venture creation and early-stage investment, where uncertainty and bounded rationality dominate.

Recent advances in large language models (LLMs) have opened new avenues for constructing synthetic agents capable of generating contextually grounded, text-based decisions. Unlike traditional ABM agents, LLM-driven personas can integrate knowledge from diverse training corpora and produce responses that resemble interview transcripts, strategic reflections, and negotiation stances \citep{Park2023, Horton2023, Argyle2023}. Early studies have begun treating LLMs as social simulators - demonstrating, for example, their capacity to model deliberation, conformity, or even cultural bias. However, systematic applications of such agents to entrepreneurship and innovation ecosystems remain rare. This paper addresses that gap by evaluating how synthetic founders and investors reason about the adoption of AI in market validation.

The empirical basis for this study comes from a prior qualitative investigation with 15 human founders, who reflected on their hopes and fears about AI-powered market validation. We conduct a comparative validation, or methodological docking experiment, using 35 LLM-driven users representing both founders and investors, treating them as computational actors within a controlled simulation environment. Our objective is not to test predictive accuracy per se, but to assess whether synthetic agents produce emergent patterns of reasoning comparable to human data, and where they diverge in systematic ways. In doing so, we position synthetic users as a methodological bridge: more expressive than rule-based ABM, yet more scalable and controllable than human fieldwork, consistent with calls for richer simulation grounded in empirical comparison \citep{Epstein2008}.

This paper contributes to the field of social simulation in three ways. First, it demonstrates a novel use of language-based synthetic personas as simulation agents, expanding the methodological repertoire beyond traditional rule-based modeling. Second, it introduces a comparative framework for evaluating how synthetic outputs relate to human-derived data, distinguishing convergent themes, partial overlaps, and insights that appear only in human or only in synthetic agents. This framework enables researchers to assess simulation fidelity while also identifying systematic divergences and blind spots. Third, the paper outlines a roadmap for applying synthetic simulations to other high-uncertainty domains, such as innovation ecosystems, organizational learning, and technology adoption. Taken together, these contributions suggest that LLM-driven personas represent a new hybrid category of social simulation - merging computational scalability with the interpretive depth of qualitative research. This study should therefore be understood as a comparative validation study, or methodological docking experiment \citep{Axtell1996}, that systematically aligns human-subject data with synthetic agents to evaluate correspondence, divergence, and blind spots. Such comparative validation directly engages with ongoing debates on model credibility and methodological rigor in computational social science.

\section{Literature Review}
\subsection{Social Simulation and Agent-Based Modeling}

Social simulation has long sought to capture the dynamics of human decision-making under conditions of complexity and uncertainty \citep{GilbertTroitzsch2005, Epstein1999}. Agent-based modeling (ABM), in particular, has provided a robust framework for representing heterogeneous actors, their decision rules, and the emergent outcomes of their interactions \citep{EpsteinAxtell1996, MacyWiller2002}. Classic applications have ranged from modeling market adoption curves \citep{Bass1969} to exploring organizational learning, diffusion of innovation, and policy interventions \citep{Fioretti2013}. While ABM excels at identifying emergent system-level patterns from simple micro-level rules, its reliance on predefined behavioral assumptions limits its ability to reflect the nuance and unpredictability of real-world human reasoning \citep{ContePaolucci2014}. This recognition echoes \citet{Epstein2008}'s call to view models not as predictive engines but as heuristic tools for generating explanation and insight.

\subsection{Beyond Rule-Based Agents: Toward Richer Simulations}

Efforts to extend ABM have introduced cognitive architectures, affective states, and empirically derived heuristics \citep{JanssenOstrom2006, Sun2006}. These enhance realism but often at the expense of scalability and transparency \citep{Grimm2006}. Moreover, such approaches still rely on explicitly coded rules, leaving them vulnerable to oversimplification and researcher bias \citep{Edmonds2005}. This tension - between parsimony and realism - has been a recurring theme in simulation research debates \citep{Edmonds2005, Epstein2008}. Complementary calls to ground simulation in psychological theory and social influence processes have also been advanced \citep{Flache2011, Jager2017}. Calls for models that are both rigorous and empirically grounded have fueled interest in generative and data-driven approaches that can complement or even transform traditional simulation paradigms \citep{Helbing2012}.

\subsection{Large Language Models as Social Simulators}

Recent advances in large language models (LLMs) provide such a pathway. The integration of LLMs into agent-based modeling has been identified as a promising new frontier in simulation, offering a way to enhance the capabilities and realism of traditional models \citep{Gao2024}. Trained on vast corpora of human language, LLMs can generate responses that approximate dialogue, strategic reasoning, and context-sensitive decision-making \citep{OpenAI2023}. Early experiments have treated LLMs as social simulators, demonstrating their capacity to simulate negotiation, group deliberation, and emergent cultural norms \citep{Park2023, Horton2023, Argyle2023}. The ``Generative Agents'' experiments showed how LLM-driven avatars could form routines, build memories, and exhibit socially coherent behavior over time \citep{Park2023}. 

Other studies highlighted LLMs’ ability to replicate survey responses, capture cognitive biases, or act as stand-ins for participants in experimental economics \citep{Argyle2023, Horton2023}. While promising, much of this work has focused on generic social interaction rather than domain-specific contexts such as entrepreneurship, innovation, or investment decision-making. The application of AI-generated synthetic users to startup market validation is a more recent innovation, leveraging large language models (LLMs) and behavioral datasets to simulate customer decision-making under realistic market scenarios \citep{Teutloff2025}, generating outputs that serve as validation evidence for investors.

\subsection{Research Gap}

Despite growing enthusiasm for LLMs as engines of social simulation, two gaps remain. First, few studies systematically compare LLM-generated agent data with human-derived data to evaluate convergence, partial overlaps, and systematic differences unique to humans or synthetics.

Second, applications to entrepreneurial ecosystems remain scarce, even though these contexts are marked by uncertainty, bounded rationality, and social signaling \citep{McMullenShepherd2006, Simon1991}. 

This paper addresses both gaps by treating AI-generated founder and investor personas as simulation agents and systematically aligning them with a prior human interview study. Addressing these gaps also requires transparency and standardized reporting, as emphasized in the ODD protocol \citep{Grimm2006}. Our study adapts this ethos by defining explicit evaluation criteria for comparing human and synthetic agents, staging a docking experiment that tests convergent themes, partial overlaps, and themes unique to human or synthetic agents.

\section{Methodology}
The model description follows the ODD (Overview, Design concepts, and Details) protocol for describing individual- and agent-based models \citep{Grimm2006, Grimm2020}.

\subsection{Overview}
This study was designed as a comparative social simulation experiment - a methodological docking exercise - to assess the credibility of AI-generated personas as simulation agents. The aim was to evaluate convergent themes (where synthetic agents reproduced human insights), partial overlaps (where related but differently framed patterns appeared), and themes unique to humans or to synthetic agents (revealing blind spots in either dataset).

The entities modeled were: human founders (baseline sample, N = 15), providing qualitative interview data; and synthetic users (N = 35), composed of founder personas configured by venture stage, resource constraints, and entrepreneurial experience, alongside investor personas configured by investment thesis, stage preference, and risk tolerance. The temporal scale was single-session interviews, and the unit of analysis was coded transcript data.
 
The experiment proceeded in two phases: (1) baseline interviews with human founders and (2) simulated interviews with synthetic users, exposed to a protocol that mirrored the human study in scope and direction while adapting wording for machine–human interaction. Outputs from both phases were analyzed thematically and compared.

\subsection{Design Concepts}
The design follows a docking experiment \citep{Axtell1996, Edmonds2005}, aligning human reasoning and synthetic reasoning to test for correspondence and identify boundaries of validity.

Themes such as ``skin in the game'' signals emerged inductively from both human and synthetic transcripts. Where themes appeared in both datasets, they were classified as convergent; where they appeared in only one dataset, they were treated as dataset-specific properties. For example, emergent logics generated only by synthetic personas - including amplified false positives, trauma blind spots, and the role of AI as a psychological buffer - were coded as synthetic-only.

Synthetic agents did not adapt dynamically across runs. Their objective was to respond to interview prompts as role-conditioned founders or investors, based on their initialization.
 
Human founders drew on lived entrepreneurial experience. Synthetic agents relied on pretrained model knowledge but did not learn across episodes.

Agents ``sensed'' interview prompts and responded as if in one-on-one interviews. No peer-to-peer interactions were modeled.

Response variability was managed through the architecture of the SyntheticUsers.com platform, which employs an ensemble-style routing agent to dynamically shuffle between multiple large language models (e.g., GPT, Claude, Llama series). This design is intended to enhance behavioral realism and mitigate single-model bias, producing diverse yet comparable outputs without manual tuning of temperature or other sampling parameters. Personas were further grounded in established personality frameworks and affective modeling, which ensured variation in responses while maintaining internal consistency across scenarios.

No collective dynamics were modeled. Results reflect aggregated patterns across multiple independent agent instantiations: 15 human founders, 30 synthetic founders, and 5 synthetic investors.

Outputs consisted of text transcripts. Responses were coded into higher-order categories using thematic analysis \citep{Braun2006}.

\subsection{Details}
Human participants were recruited and interviewed under IRB-approved procedures in December 2024 - January 2025. Synthetic agents were generated in July 2025 through SyntheticUsers.com, a commercial platform that integrates outputs from a conglomerate of large language models (LLMs). The platform employs an ensemble-style routing agent that dynamically shuffles between models (e.g., GPT, Claude, Llama series), enhancing behavioral realism and mitigating single-model bias. To generate human-like responses, personas are grounded in established personality frameworks \citep{McCraeCosta1997} and supplemented with affective modeling to ensure varied and realistic reactions to market stimuli. A Retrieval-Augmented Generation (RAG) layer incorporates domain-specific knowledge (e.g., market reports, academic articles), and personas are fine-tuned against large-scale behavioral datasets such as the U.S. General Social Survey to align them with demographic and psychographic profiles \citep{SyntheticUsers2025}.

For this study, the synthetic cohort comprised thirty founder personas representing diverse archetypes - technical founders with AI/ML expertise, serial entrepreneurs with prior validation experience, non-technical co-founders, and first-time founders in resource-constrained environments - alongside five early-stage investor personas with a focus on AI and venture innovation. These synthetic users were not intended to represent actual individuals but to elicit structured, comparable perspectives that could be systematically contrasted with the human interview data. This design allowed decision-making behaviors to be simulated in a way that was comparable to Phase 1 participants, while offering scalability and consistency at a fraction of the cost and time.

The credibility of the synthetic agents was assessed using three criteria adapted from simulation science: face validity \citep{Troitzsch2004}, construct fidelity \citep{Sun2006}, and dataset-specific emergence \citep{Helbing2012}, referring to patterns that appear only in human or only in synthetic agents.

While human interview transcripts cannot be shared due to IRB confidentiality constraints, the materials required to reproduce the synthetic phase of this study have been archived with the authors and can be shared upon request, consistent with open-science best practices.

\section{Results}
This study identified four categories of outcomes: convergent themes (shared across humans and synthetic personas), partial overlaps (similar concerns framed differently), human-only themes, and synthetic-only themes. Together, these patterns illustrate both the fidelity and limitations of LLM-driven social simulation in entrepreneurial contexts.

\subsection{Convergent Themes: Evidence of Simulation Fidelity}
Several themes appeared consistently across both human founders and synthetic personas, suggesting that the agents captured core decision heuristics relevant to early-stage validation.

\medskip 
\textbf{Commitment-based signals}
\begin{itemize}
    \item \textbf{Human founder:} ``[Only] if people are paying money, then it's proven.''
    \item \textbf{Synthetic founder personas:} Emphasized that validation only ``counts'' when customers show behavioral commitment---pre-orders, deposits, or onboarding effort---not just verbal interest.
\end{itemize}

\medskip
\textbf{The ‘black box’ trust barrier}
\begin{itemize}
    \item \textbf{Human founder:} ``That is the question of this century. How can I prove that this information is true?''
    \item \textbf{Synthetic personas:} Require clear, plain-language explanations of the methodology, data sources, and bias mitigation strategies to build trust.
\end{itemize}

\medskip
\textbf{Efficiency gains}
\begin{itemize}
    \item \textbf{Human founder:} ``Things that took six months now take six hours.''
    \item \textbf{Synthetic personas:} Stated that AI validation speeds up the due diligence process and reduces costs, giving startups a competitive advantage when time is critical.
\end{itemize}

\medskip 
These alignments suggest that synthetic personas can replicate domain-relevant reasoning, lending credibility to their use as simulation agents.

\subsection{Partial Overlaps: Differently Framed Concerns}
Some themes emerged in both datasets but were framed differently.

\medskip 
\textbf{Edge-case sensitivity}
\begin{itemize}
    \item \textbf{Human founders:} Worried that averages might obscure strategically decisive outliers---for example, a few B2B contracts or niche user groups whose adoption (or rejection) could determine overall viability.
    \item \textbf{Synthetic personas:} Warned that AI may miss irrational sticky points, long-tail behavior, or unmodellable nuances.
\end{itemize}

\medskip
\textbf{AI as a psychological buffer}
\begin{itemize}
    \item \textbf{Human founder:} ``The most intimidating thing is you go out and have to talk to these people.''
    \item \textbf{Synthetic personas:} Stated AI offers the allure of ``clean,'' objective data that can be used to manage internal anxieties or external investor expectations.
\end{itemize}

\medskip 
These overlaps reveal complementary lenses: humans stress outlier contracts and emotional discomfort, while synthetics articulate blind spots and coping strategies.

\subsection{Human-Only Themes: Lived Experience and Relational Capital}
Some themes appeared only in the human interviews, underscoring the irreplaceable role of lived experience and relational trust.

\medskip 
\textbf{Relational \& advocacy value}
\begin{itemize}
    \item \textbf{Human founder:} ``Not only they will be your first adopters, they're going to be your evangelists...and even occasionally financially support you.''
\end{itemize}

\medskip
\textbf{Moonshot skepticism}
\begin{itemize}
    \item \textbf{Human founders:} Argued that while AI might assist with incremental or well-defined markets, it would struggle to anticipate disruptive opportunities like Airbnb or SpaceX.
\end{itemize}

\medskip 
These themes reflect that early-stage entrepreneurship often relies on social capital and skepticism about radical novelty---factors that current synthetic agents cannot model.

\subsection{Synthetic-Only Themes: Emergent Logics of AI Agents}
Synthetic personas also generated distinctive reasoning patterns not observed among human founders, reflecting the novel affordances and biases of LLM-driven agents.

\medskip 
\textbf{Amplified false positives}
\begin{itemize}
    \item \textbf{Synthetic personas:} Expressed fear that synthetic users, if not meticulously designed, might inadvertently amplify ``polite lies'' – one of the main drawbacks of traditional market research.
\end{itemize}

\medskip
\textbf{Experiential trauma blind spots}
\begin{itemize}
    \item \textbf{Synthetic personas:} Warned that markets scarred by past failures may resist new offerings, a barrier AI models may fail to detect.
\end{itemize}

\medskip 
These synthetic-only perspectives highlight both the risks and potential extensions of LLM-based agents, surfacing concerns that humans did not articulate.

\subsection{Synthesis}
Taken together, the findings indicate that LLM-driven personas can operate as semi-faithful simulation agents---capable of reproducing certain human heuristics, partially overlapping with others, generating novel synthetic-only logics, and omitting salient human-only concerns.

To make these contrasts transparent, Table~\ref{tab:comparative_themes} summarizes the core themes across human founders and synthetic personas. This comparative view highlights how convergences strengthen claims of face and construct validity, partial overlaps reveal complementary framings, and human-only or synthetic-only themes clarify both blind spots and opportunities.

\begin{table}[htbp]
\centering
\caption{Comparative themes across human founders and synthetic personas}
\label{tab:comparative_themes}
\begin{tabularx}{\textwidth}{
    >{\raggedright\arraybackslash}p{2.5cm} 
    >{\raggedright\arraybackslash}p{3cm} 
    X 
    X
}
\toprule
\textbf{Category} & \textbf{Specific Theme} & \textbf{Human Founders} & \textbf{Synthetic Personas} \\
\midrule
\multirow{3}{*}{Convergent} & Commitment Signals & ``Skin in the game'' (payment, effort). & Reproduced emphasis on commitment. \\
\cmidrule{2-4}
 & Black-box Trust Barrier & Called for transparency/auditability. & Echoed concerns, added detail. \\
\cmidrule{2-4}
 & Efficiency Gains & Stressed speed as a competitive edge. & Highlighted portfolio-level efficiency. \\
\midrule
\multirow{2}{*}{Partial Overlap} & Edge-case Sensitivity & Worried about smoothing out outliers. & Flagged irrational sticky points, long-tail behaviors. \\
\cmidrule{2-4}
 & AI as Psychological Buffer & Described interviews as stressful. & Explicitly framed AI as coping tool. \\
\midrule
\multirow{2}{*}{Human-Only} & Relational \& Advocacy Value & Valued early customer relationships and advocacy. & Not present. \\
\cmidrule{2-4}
 & Moonshot Skepticism & Questioned AI’s ability to test radical markets. & Not present. \\
\midrule
\multirow{2}{*}{Synthetic-Only} & Amplified False Positives & Not mentioned. & Warned of ``polite lies'' inflating optimism. \\
\cmidrule{2-4}
 & Experiential Trauma Blind Spot & Not mentioned. & Warned AI may miss resistance from past failures. \\
\bottomrule
\end{tabularx}
\end{table}

The comparative table underscores that fidelity was evident in convergent heuristics such as commitment signals, efficiency gains, and black-box trust concerns. Partial overlaps revealed how similar issues were framed differently---human founders stressing outlier contracts or the stress of customer interviews, while synthetic personas emphasized irrational blind spots and the buffering role of AI. Human-only themes highlighted relational and advocacy value as well as skepticism toward radical ``moonshot'' ideas, whereas synthetic-only themes surfaced risks such as amplified false positives and the overlooking of experiential trauma.

For simulation research, these distinctions are not merely descriptive but methodological. Convergent findings strengthen claims of face and construct validity, partial overlaps point to complementary perspectives worth further investigation, and human-only and synthetic-only themes delineate systematic blind spots. Taken together, they reinforce the value of docking experiments as a structured approach to evaluating new classes of agents against empirical benchmarks, clarifying both their explanatory potential and their boundaries of realism.

\section{Discussion}
\subsection{Synthetic Personas as a Hybrid Simulation Approach}

Our findings show that LLM-driven personas can reproduce core entrepreneurial heuristics (e.g., ``skin in the game'' signals, distrust of black-box methods) while also generating novel reasoning patterns. This places synthetic personas in a hybrid simulation category: richer and more linguistically expressive than traditional ABM, yet more scalable and controllable than field-based qualitative studies \citep{ContePaolucci2014, Edmonds2005}.

For simulation research debates on parsimony versus realism, this suggests a new middle ground---agents that are not purely rule-based, but also not ``free text'' without structure.

Importantly, this hybrid category also speaks to the long-running question of cognitive realism in social simulation. LLM personas extend beyond parsimony by producing text that exhibits psychological nuance, bias, and strategic reflection. At the same time, their limitations---particularly the absence of lived history, relational trust, and emotional consequence---mark the boundaries of this realism. Thus, our results show both the promise and the current frontier of cognitive realism in agent design.

\subsection{Assessing Fidelity, Divergence, and Blind Spots}

We proposed a comparative docking experiment to evaluate four outcome types: convergent themes, partial overlaps, human-only themes, and synthetic-only themes. Table~\ref{tab:comparative_themes} illustrates these categories directly, contrasting human and synthetic findings in a compact comparative format.

The convergent themes in Table~\ref{tab:comparative_themes} (e.g., commitment signals, black-box concerns, efficiency) confirm that synthetic personas can reproduce key heuristics, providing evidence of face and construct validity.

The partial overlaps (e.g., edge-case sensitivity, psychological buffering) reveal issues that surfaced in both datasets but were framed differently, suggesting complementary perspectives rather than error. Synthetic-only themes (e.g., amplified false positives, trauma blind spots) demonstrate novel but coherent logics, while human-only themes (e.g., relational advocacy, moonshot skepticism) highlight concerns absent in synthetic outputs. These distinctions are not errors but systematic differences that can serve as hypotheses for further inquiry, consistent with \citet{Epstein2008}'s argument that models should be valued for the questions they provoke.

These patterns highlight the cognitive limits of LLM-based personas. As Table~\ref{tab:comparative_themes} indicates, human-only insights cluster around lived experience, relational trust, and skepticism toward radical novelty---factors outside the reach of current models. Synthetic-only insights, by contrast, reveal potential risks of AI-driven amplification and neglect of negative history. Docking experiments thus help delineate not only where synthetic agents succeed but also where their realism stops.

\subsection{Implications for Simulation Science}

For the wider simulation community, three implications emerge:

\begin{description}
    \item[Methodological innovation] --- Language-based agents extend ABM by offering contextually rich, psychologically plausible responses.
    
    \item[Comparative validation] --- Aligning human and synthetic data provides a structured method to test correspondence and limits, echoing calls for standardized reporting and transparency \citep{Grimm2006, Edmonds2005}. Our comparative docking also aligns with ongoing concerns in computational social science about how models can and should be validated against empirical data \citep{Squazzoni2020}, reinforcing that methodological rigor must be assessed not only in parsimony but also in empirical correspondence.
    
    \item[Hypothesis generation] --- Dataset-specific themes highlight where simulation might reveal unexplored mechanisms, which can then be tested empirically.
\end{description}

Together, these contributions position docking experiments as a systematic way to evaluate the credibility of new agent types.

\subsection{Limitations and Ethical Considerations}

The study also highlights limitations. Synthetic agents cannot embody lived history, experiential trauma, or relational trust. Outputs are constrained by model training data and prompt design, and may amplify embedded biases \citep{Bender2021}. As \citet{Weidinger2021} note, language models also pose systemic ethical risks, from representational harms to misuse potential. For social simulation, these concerns are not peripheral but central to validation debates: model credibility must extend beyond technical accuracy to include transparency, accountability, and responsible use. Researchers should therefore embed ethics directly into evaluation criteria, ensuring that simulation design choices (e.g., temperature settings, role prompts) are documented and reproducible \citep{Grimm2006}.

\subsection{Future Directions}

We see three promising pathways:
\begin{itemize}
    \item \textbf{Benchmarking:} Systematically comparing synthetic outputs with larger human datasets to quantify fidelity and boundary conditions.
    
    \item \textbf{Hybrid simulation designs:} Combining rule-based ABM with language-based agents to integrate parsimony, scalability, and linguistic realism.
    
    \item \textbf{Cross-domain application:} Extending docking experiments to domains such as organizational learning, policy analysis, or innovation diffusion, while maintaining rigorous validation protocols.
\end{itemize}

\section{Conclusion}

This study has demonstrated how large language models can be used to generate synthetic personas as simulation agents. By staging a comparative validation study---a methodological docking experiment---we systematically aligned human founder interviews with synthetic founder and investor personas. The results showed a four-part pattern: convergent themes where synthetic agents reproduced core heuristics, partial overlaps where similar issues were framed differently, human-only themes highlighting lived experience and relational capital, and synthetic-only themes surfacing risks such as amplified false positives and trauma blind spots.

For computational social science, the contribution lies in three areas. First, we establish comparative docking as a methodological framework for evaluating new classes of agents against empirical data, offering a structured way to assess credibility. Second, we extend the representational scope of simulation beyond rule-based agents by showing that language-based personas can yield linguistically rich, psychologically plausible decision traces. Third, we highlight dataset-specific themes as a resource for hypothesis generation rather than model failure, consistent with long-standing arguments that models should be valued for the questions they provoke \citep{Epstein2008}.

At the same time, the study underscores limitations: synthetic personas cannot substitute for lived experience, are shaped by training biases, and carry ethical risks if deployed without transparency. Their most appropriate use is within hybrid simulation designs, where synthetic outputs accelerate exploration and human data provide grounding.

Looking forward, the development of benchmarks, hybrid architectures, and cross-domain docking experiments can help consolidate this approach. The promise of synthetic personas is not that they replace traditional ABM or fieldwork, but that they expand the toolkit of social simulation---offering scalable yet empirically sensitive instruments for probing decision-making under uncertainty.

\section*{Data Availability}
The datasets generated and analyzed during the current study are available from the corresponding author on reasonable request. This includes the full, non-sensitive dataset of responses from the synthetic persona replication study and the anonymized transcripts from the human-founder interviews, the latter of which are not publicly available to protect participant confidentiality.

\bibliographystyle{apalike}
\bibliography{references}

\end{document}